\documentclass[11pt,a4paper]{article}
\usepackage[utf8]{inputenc}
\usepackage[T1]{fontenc}
\usepackage[english]{babel}
\usepackage{newtxtext} % Times Roman clone
\usepackage{newtxmath} % Math fonts to match
\usepackage{microtype} % Improves justification and spacing
\usepackage{csquotes} % Context-sensitive quotations
\usepackage{amsmath}
\usepackage{amssymb}
\usepackage{graphicx}
\usepackage{booktabs} % Professional table formatting
\usepackage{geometry}
\usepackage{authblk} % For author affiliations
\usepackage{enumitem} % Control over list environments
\usepackage{listings} % Code listings
\usepackage{abstract}
\usepackage{float} % Better float placement control
\usepackage[hidelinks, colorlinks=true, linkcolor=blue!50!black, citecolor=green!50!black, urlcolor=magenta!50!black]{hyperref}
\usepackage{xurl}
\usepackage[figurename=Fig., tablename=Tab., labelfont=bf, textfont=it]{caption}
\usepackage{xcolor}
\definecolor{codegreen}{rgb}{0,0.6,0}
\definecolor{codegray}{rgb}{0.5,0.5,0.5}
\definecolor{codepurple}{rgb}{0.58,0,0.82}
\definecolor{backcolour}{rgb}{0.98,0.98,0.98}
\lstdefinestyle{jsonstyle}{
    backgroundcolor=\color{backcolour},
    commentstyle=\color{codegreen},
    keywordstyle=\color{blue},
    numberstyle=\tiny\color{codegray},
    stringstyle=\color{codepurple},
    basicstyle=\ttfamily\footnotesize,
    breakatwhitespace=false,
    breaklines=true,
    captionpos=b,
    keepspaces=true,
    numbers=left,
    numbersep=5pt,
    showspaces=false,
    showstringspaces=false,
    showtabs=false,
    tabsize=2,
    morestring=[b]",
    literate=
     *{0}{{{\color{black}0}}}{1}
      {1}{{{\color{black}1}}}{1}
      {2}{{{\color{black}2}}}{1}
      {3}{{{\color{black}3}}}{1}
      {4}{{{\color{black}4}}}{1}
      {5}{{{\color{black}5}}}{1}
      {6}{{{\color{black}6}}}{1}
      {7}{{{\color{black}7}}}{1}
      {8}{{{\color{black}8}}}{1}
      {9}{{{\color{black}9}}}{1}
}
\lstdefinelanguage{json}{
    morestring=[b]",
    morecomment=[l]{//},
    morekeywords={true,false,null},
    sensitive=false,
    alsoletter={:},
    moredelim=[l][\color{black}\bfseries]{"},
}
\providecommand{\keywords}[1]
{
  \small
  \textbf{\textit{Keywords---}} #1
}
\title{\Large\textbf{Verification of Lightning Network Channel Balances with Trusted Execution Environments (TEE)}}
\author[1]{Vikash Singh}
\author[2]{Barrett Little}
\author[4]{Philip Hayes}
\author[4]{Max Fang}
\author[5]{Matthew Khanzadeh}
\author[1]{Alyse Killeen}
\author[3]{Sam Abbassi}
\affil[1]{Stillmark}
\affil[2]{IBEX Mercado}
\affil[3]{Hoseki}
\affil[4]{Lexe}
\affil[5]{Independent}
\setlist{itemsep=0pt, topsep=3pt, partopsep=0pt}
\begin{document}
\date{December 11, 2025}
\maketitle
\begin{abstract}
Verifying the private liquidity state of Lightning Network (LN) channels is desirable for auditors, service providers, and network participants who need assurance of financial capacity. Current methods often lack robustness against a malicious or compromised node operator. This paper introduces a methodology for the verification of LN channel balances. The core contribution is a framework that combines Trusted Execution Environments (TEEs) with Zero-Knowledge Transport Layer Security (zkTLS) to provide strong, hardware-backed guarantees. In our proposed method, the node's balance-reporting software runs within a TEE, which generates a remote attestation quote proving the software's integrity. This attestation is then served via an Application Programming Interface (API), and zkTLS is used to prove the authenticity of its delivery. We also analyze an alternative variant where the TEE signs the report directly without zkTLS, discussing the trade-offs between transport-layer verification and direct enclave signing. We further refine this by distinguishing between \enquote{Hot Proofs} (verifiable claims via TEEs) and \enquote{Cold Proofs} (on-chain settlement), and discuss critical security considerations including hardware vulnerabilities, privacy leakage to third-party APIs, and the performance overhead of enclaved operations.
\end{abstract}
\section{Introduction}
\label{sec:introduction}
The Lightning Network (LN) \cite{PoonDryja2016} operates as a vital layer-2 protocol for scaling Bitcoin \cite{Nakamoto2008} transactions. Its efficiency stems from the use of off-chain payment channels where liquidity—the distribution of funds between channel partners—is inherently private. This privacy is a double-edged sword: while beneficial for users, it creates significant challenges for external entities, such as auditors, liquidity marketplaces, or counterparties who require verifiable proof of a node's channel reserves. Establishing such proofs in a trust-minimized way is essential for risk assessment, financial accountability, and the maturation of the LN ecosystem.

This work extends prior TEE-based LN systems such as Teechan \cite{Lind2016Teechan} and Teechain \cite{Lind2017Teechain} by focusing on verifiable balance proofs rather than channel establishment.

This paper explores the spectrum of methodologies for verifying LN channel balances, culminating in a proposal for a novel, robust framework. We frame the problem through a progression of trust models:

\begin{enumerate}[label=\arabic*)]
    \item \textbf{External Observation:} At the baseline, an external party can attempt to infer liquidity through network-level actions like payment probing. This approach requires no cooperation from the target node but offers limited precision and carries privacy implications.
    \item \textbf{Software-Level Attestation:} As an improvement, a node can report its own balance via an API, with zkTLS providing cryptographic proof that the node's API endpoint indeed served that specific report. This proves what the node \textit{claimed}, but fundamentally relies on trusting the integrity of the node's software stack.
    \item \textbf{Hardware-Attested Computation:} To address the fundamental limitation of software integrity, this paper's primary contribution is a novel, robust methodology that combines TEEs with zkTLS. By running the balance-reporting logic inside a TEE, the node operator can provide hardware-backed proof (a remote attestation quote) that a specific, audited version of the software was executed. zkTLS then ensures the authentic delivery of this high-assurance attestation.
\end{enumerate}

Furthermore, we posit that robust verification requires two distinct classes of proofs. We introduce the concept of \textbf{\enquote{Hot Proofs}}—verifiable claims generated by TEEs for speed and operational continuity—and \textbf{\enquote{Cold Proofs}}—the ultimate settlement of funds on-chain, which serves as the final arbiter. The principal contribution of this work is the design and analysis of the \enquote{Hot Proof} TEE-enhanced verification framework, presented within a comparative overview that establishes its position as a state-of-the-art solution for robust channel balance verification.

The paper is structured as follows: Section~\ref{sec:preliminaries} reviews foundational concepts. Section~\ref{sec:spectrum} details the first two verification methodologies. Section~\ref{sec:proposed_framework} presents our proposed TEE-based framework in detail, including necessary protections against stale state attacks. Section~\ref{sec:comparative_analysis} provides a comparative discussion of all three methods. Section~\ref{sec:industrial_relevance} outlines the commercial and industrial implications of this technology. Section~\ref{sec:limitations_future_work} outlines broader limitations and future work, and Section~\ref{sec:conclusion} concludes the paper.

\section{Background and Preliminaries}
\label{sec:preliminaries}

\subsection{Lightning Network (LN)}
\label{ssec:prelim_ln}
The LN consists of bidirectional payment channels established between pairs of nodes via on-chain funding transactions. Within a channel, parties make numerous off-chain payments by updating commitment transactions that reflect the current distribution of funds (local and remote balances). Outbound liquidity for a node in a channel is its local balance available for sending payments, minus channel reserves and committed HTLCs.

\subsection{Payment Probing}
\label{ssec:prelim_probing}
Payment probing involves sending HTLCs (often with unfulfillable preimages) to test path viability. A successful probe of amount $X$ suggests liquidity $\ge X$. A failure with an error like \texttt{temporary\_channel\_failure} at a specific hop suggests liquidity $< X$ on that hop's outgoing channel. Probing can estimate liquidity bounds but has privacy drawbacks \cite{Tikhomirov2020probing}.

\subsection{zkTLS and TLSNotary}
\label{ssec:prelim_zktls}
Zero-Knowledge Transport Layer Security (zkTLS) and related HTTPS attestation schemes enable a Prover to prove to a Verifier that specific data was part of an encrypted TLS session with a particular Server, without revealing the session's master key or other sensitive session data. A prominent protocol suite for achieving this is \textbf{TLSNotary} \cite{TLSNotary}. It operates on a two-party trust model involving the data user (the \textit{Prover}) and an independent third party (the \textit{Notary}). The core mechanism relies on \textbf{Multi-Party Computation (MPC)} to prevent unilateral forgery by the Prover. At a high level, the process is as follows:

\begin{enumerate}
    \item The Prover initiates a TLS connection to the target Server (e.g., their LND node's API).
    \item Simultaneously, the Prover engages in an MPC protocol with the Notary.
    \item During the TLS handshake between the Prover and the Server, the crucial TLS master secret (from which session keys are derived) is not fully known by either the Prover or the Notary. Instead, it is generated as a shared secret between them through their MPC interaction.
    \item When the Server sends its encrypted response (e.g., a JSON balance report), the Prover and Notary jointly use their secret shares to decrypt only the specific parts of the response that need to be attested to. The Notary does not learn the full session content, preserving the Prover's privacy.
    \item The Notary, having participated in a successful decryption, can then issue a signed statement (an attestation) confirming that the revealed plaintext was indeed part of that valid TLS session.
\end{enumerate}

The final proof given to an Auditor combines transcript information from the Prover and the signed attestation from the Notary, providing strong evidence of the data's authenticity and origin from the server.

\subsection{Trusted Execution Environments (TEEs)}
\label{ssec:prelim_tee}
TEEs, such as Intel SGX and AMD SEV, are hardware-isolated environments protecting code and data from the host system, including the OS. A key feature is \textbf{remote attestation}. TEE hardware can generate a cryptographic quote signed by a hardware-private key. This quote includes a measurement (hash) of the code loaded into the TEE enclave (e.g., MRENCLAVE), identifying the software version. It can also include user-defined \enquote{report data} (e.g., a hash of the application's output), binding the attestation to a specific result.

\section{A Spectrum of Verification Methodologies}
\label{sec:spectrum}
This section analyzes two foundational approaches to LN reserve verification, setting the stage for our primary proposal.

\subsection{Method 1: External Observation via Auditor-Driven Probing}
\label{ssec:method_probing}
\textbf{Mechanism:} The Auditor uses their own LN node to actively send payment probes through or to the target node's channels. By observing successes and failures of probes of varying amounts, the Auditor iteratively refines an estimate of the available outbound liquidity in a target channel segment.

\textbf{Analysis:} This method is advantageous as it is entirely external and requires no active cooperation from the target node for public channels. However, it is imprecise, providing only liquidity bounds. It is also invasive, consuming network resources and potentially revealing information about the probe path to other parties. It does not produce a cryptographic proof from the target and can be thwarted by anti-probing measures.

\subsection{Method 2: Software-Level Attestation via zkTLS}
\label{ssec:method_zktls}
\textbf{Mechanism:} The target node operator (Prover) uses their own node to report its channel balance(s) via an API. They then use a zkTLS protocol like TLSNotary to prove to the Auditor (Verifier) that this report genuinely originated from their node's API. A concrete example involves using the standard API for LND (Lightning Network Daemon) \cite{LNDAPI}. While this paper primarily uses LND for demonstration, the methodology is equally applicable to other implementations such as Core Lightning (CLN) or custom nodes built with the Lightning Dev Kit (LDK), provided they expose verifiable state interfaces.

The Prover's LND node exposes a gRPC or REST API endpoint, such as the \texttt{ChannelBalance} RPC, available at the REST path \texttt{/v1/balance/channels}. This endpoint returns a JSON object with fields detailing the node's aggregate channel balances. The Prover queries this endpoint on their own node and generates a zkTLS proof for the response, an example of which is shown in Listing~\ref{lst:lnd_api_response}.

\begin{lstlisting}[language=json, caption={Example API Response from LND's `/v1/balance/channels` Endpoint}, label=lst:lnd_api_response]
{
  "local_balance": {
    "sat": "1234567",
    "msat": "1234567000"
  },
  "remote_balance": {
    "sat": "765433",
    "msat": "765433000"
  },
  "unsettled_local_balance": { "sat": "0", "msat": "0" },
  "unsettled_remote_balance": { "sat": "0", "msat": "0" },
  "pending_open_local_balance": { "sat": "0", "msat": "0" },
  "pending_open_remote_balance": { "sat": "0", "msat": "0" }
}
\end{lstlisting}

The Prover provides this JSON response along with the TLSNotary proof to the Auditor. The Auditor verifies the proof, confirming that the Prover's LND node (identified by its TLS certificate) authentically served this balance report.

\textbf{Analysis:} This method is a significant improvement, as it produces a cryptographic attestation of what the node \textit{claims}. However, its security rests on a crucial assumption: the Auditor must trust the integrity of the LND software running on the Prover's machine. If the node operator has tampered with their own LND software to make the API lie, zkTLS will faithfully attest to this lie.

\section{Proposed Framework: Robust Verification with TEE and zkTLS}
\label{sec:proposed_framework}
To address the software integrity limitation of the zkTLS-only approach, we propose our primary contribution: a novel methodology that integrates TEEs for hardware-backed assurance.

\subsection{Motivation}
\label{ssec:tee_motivation}
The core motivation is to protect against a malicious host operator (the Prover) who might tamper with their own node software to falsify balance reports. By running the critical LND balance-reporting logic inside a TEE, the Prover can generate a TEE remote attestation quote, proving the integrity of this logic to the Auditor.

\subsection{The Complementary Roles of TEE and zkTLS}
\label{ssec:tee_zktls_roles}
It is crucial to understand why both TEE and zkTLS are necessary for a fully robust system. They solve different, complementary problems to create a layered defense against a malicious Prover.

\begin{itemize}
    \item \textbf{The TEE proves the \textit{what}:} The TEE's remote attestation quote provides a hardware-backed guarantee that a specific, untampered version of software (\textit{what}) correctly processed some internal state and produced a specific result (the balance report). It ensures the integrity of the computation.
    \item \textbf{zkTLS proves the \textit{who} and \textit{when}:} The TEE quote on its own is just a piece of static, signed data. A malicious Prover could save a valid quote from a time their balance was high and replay it later. zkTLS defeats this by attesting to the \textbf{delivery transcript}. It proves that a specific node (\textit{who}, identified by its live TLS certificate) delivered that specific TEE quote at a specific, notarized time (\textit{when}).
\end{itemize}

Analogy: The TEE quote is like a letter with an unforgeable government seal certifying its contents. The zkTLS proof is like a notarized affidavit, with today's date, confirming that a specific person handed you that sealed letter. Both are required for full confidence in our primary architecture.

\subsection{Variant: Direct TEE Attestation (No zkTLS)}
\label{ssec:no_zktls}
While our primary framework leverages zkTLS for transport authentication, a leaner architectural variant exists that relies solely on TEE primitives. In a \textbf{Direct TEE Attestation} model, the zkTLS layer is removed, and trust is placed entirely on the enclave's ability to sign data.

\textbf{Mechanism:}
\begin{enumerate}
    \item \textbf{Enclave Signing Keys:} The enclaved LND service generates a key pair $(pk_{enc}, sk_{enc})$ solely for signing reports. The public key is exported and verified as part of the initial TEE quote.
    \item \textbf{Nonce Binding (The "When"):} The Auditor provides a fresh nonce to the Prover; the Prover includes the nonce and timestamp; the Auditor verifies recency and that the nonce matches.
    \item \textbf{Execution:} The enclave executes the balance check, concatenates the result with the nonce and current timestamp, and signs it directly: $\sigma = \text{Sign}_{sk_{enc}}(\text{Balance} || N || \text{Timestamp})$.
    \item \textbf{Verification:} The Auditor verifies the signature $\sigma$ against the enclave's trusted public key.
\end{enumerate}

\textbf{Trade-offs:} This variant significantly reduces the Trusted Computing Base (TCB) by removing the MPC-based zkTLS stack. It also lowers latency by eliminating the need for MPC coordination with a Notary. However, it requires an \textit{interactive} protocol (Auditor must provide the nonce) and custom verification logic on the Auditor's side. In contrast, our primary zkTLS-based approach allows for the creation of non-interactive, portable proofs of historical API responses that can be verified using standard TLS verification tools, which is often preferable for public auditing scenarios.

\subsection{Comparison with Integrated Remote Attestation TLS (RA-TLS)}
\label{ssec:ra_tls_comparison}
It is important to distinguish the proposed TEE + zkTLS framework from another approach known as Remote Attestation TLS (RA-TLS). While both leverage TEEs, they have different architectures and address distinct goals.

\begin{itemize}
    \item \textbf{RA-TLS authenticates the endpoint at connection time.} In an RA-TLS protocol, the TEE remote attestation is integrated directly into the TLS handshake. The server's X.509 certificate is extended to include a TEE quote, which proves that the TLS private key is held within an enclave running specific, untampered software. The client's TLS library must be modified to verify this quote during the handshake. The primary result is a highly secure channel where the client has upfront assurance that it is communicating directly with trusted software. The main use case is for ongoing, confidential communication with a trusted endpoint.
    \item \textbf{The proposed framework authenticates the transcript after the fact.} Our method uses a standard TLS session to retrieve data. The TEE quote is part of the application-level payload. The zkTLS protocol (like TLSNotary) is then used separately to create a portable, non-repudiable proof that this payload was served by the specific server at a specific time. The primary result is a shareable audit record that can be verified by any third party without them needing to connect to the server themselves.
\end{itemize}

In essence, RA-TLS secures the \textit{channel}, while our proposed method uses zkTLS to secure the \textit{transcript}. For the purpose of creating verifiable audit records that can be shared and validated offline, securing the transcript is the required function, making the zkTLS approach more flexible for third-party verification scenarios.

\subsection{System Architecture and Protocol Flow (zkTLS Model)}
\label{ssec:tee_protocol}
\begin{enumerate}
    \item \textbf{TEE-Enclaved LND Service:} The critical components of the LND node software, particularly the services responsible for accessing channel state and handling the \texttt{/v1/balance/channels} API call, run inside a TEE enclave on the Prover's node, \texttt{Node\_P}. Note that while we utilize LND for this reference architecture, modular frameworks like the Lightning Dev Kit (LDK) and Core Lightning (CLN) are particularly advantageous here. They allow developers to isolate only the critical state and signing logic within the TEE while leaving heavy I/O operations on the untrusted host. The TEE hardware measures the code loaded into this enclave, resulting in a unique software measurement (e.g., MRENCLAVE).
    \item \textbf{TEE-Bound Attestation Generation:} When queried by the Prover's client tool, the enclaved LND service performs its balance lookup. It then requests the TEE hardware to generate an attestation quote (\texttt{tee\_quote}). Crucially, it includes a hash of the JSON balance report (from Listing~\ref{lst:lnd_api_response}) in the quote's \enquote{report data} field.
    \item \textbf{API Response with TEE Payload:} The node's overall API returns a package containing both the JSON balance report and the TEE attestation payload.
    \begin{lstlisting}[language=json, caption={Example TEE-Enhanced API Response Package}, label=lst:tee_package_response]
{
  "balance_report": { /* JSON object from Listing 1 */ },
  "tee_attestation_payload": {
      "quote": "BASE64...",
      "cert_chain": ["CERT1", "CERT2", "..."]
  }
}
    \end{lstlisting}
    \item \textbf{zkTLS Attestation of Delivery:} The Prover's client tool generates a zkTLS proof for this entire package being served by \texttt{Node\_P}'s API.
    \item \textbf{Submission:} The Prover submits the zkTLS proof and the plaintext package to the Auditor.
\end{enumerate}

\begin{figure}[H]
    \centering
    \includegraphics[width=0.9\textwidth]{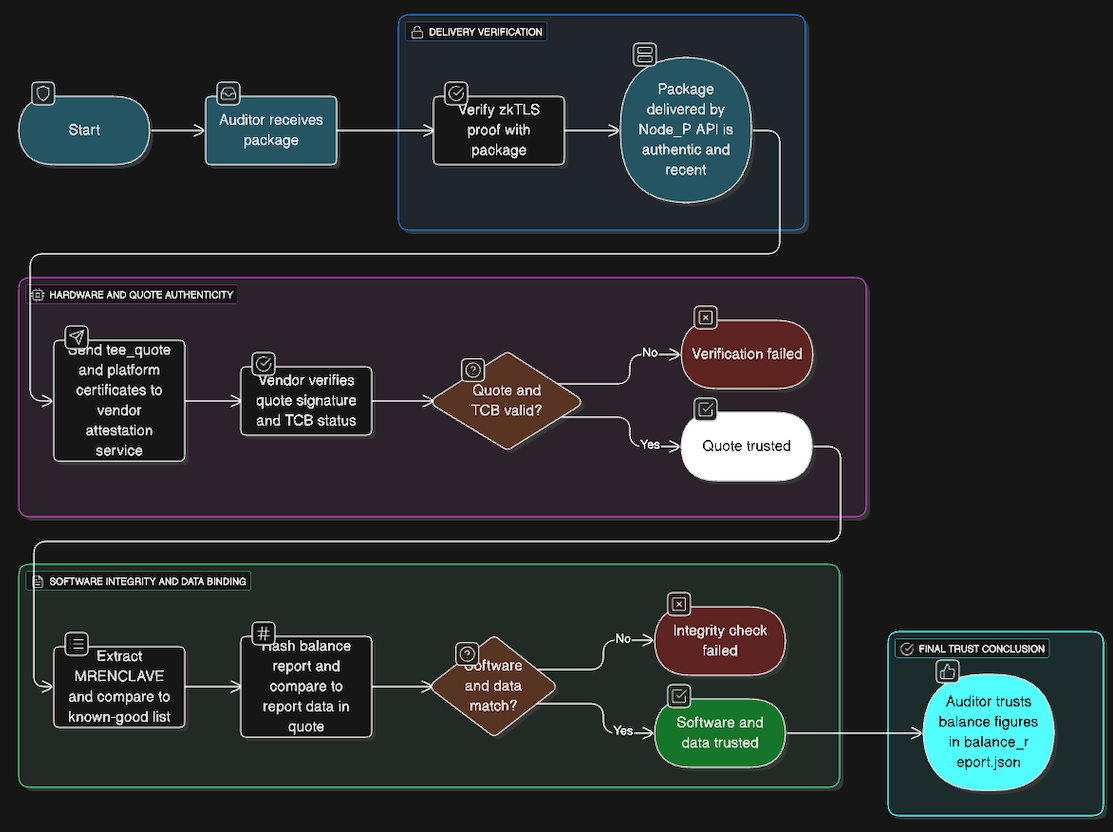}
    \caption{Architecture of the TEE + zkTLS Verification Framework. The diagram illustrates the division between the untrusted host and the secure Enclave, and how zkTLS validates the delivery of the attestation.}
    \label{fig:architecture}
\end{figure}

\subsection{Addressing Stale State via Enclaved Chain Synchronization}
\label{ssec:stale_state}
A significant challenge in TEE-based proof of reserves is the risk of \enquote{stale state} replay attacks. A malicious node operator could disconnect the TEE from the network and feed it outdated state data where a channel appears open, when in reality the channel was closed (potentially via a justice transaction).

To mitigate this, the enclave must independently verify the \enquote{freshness} of the channel state relative to the Bitcoin blockchain. We propose an \textbf{Enclaved Chain Synchronization} mechanism:

\begin{itemize}
    \item \textbf{Enclaved TLS Client:} The enclave code must include a lightweight TLS client (e.g., using libraries like \texttt{rustls}).
    \item \textbf{Trusted Roots:} The enclave must commit to a set of trusted \texttt{webpki} roots (Certificate Authorities). This ensures that when the enclave connects to an external server, the node operator cannot perform a Man-in-the-Middle (MitM) attack.
    \item \textbf{Freshness Check:} Before generating the \texttt{tee\_quote}, the enclave connects to a trustworthy third-party chain data API (e.g., Esplora or Electrum) over TLS. It fetches the current block hash and height and verifies that its channel funding transactions are still unspent in the current UTXO set. The enclave also validates that no large pending HTLCs exceed the local balance.
    \item \textbf{Liveness and Trust:} While this introduces a dependency on the third-party API, liveness risks can be mitigated. In high-stakes scenarios, the Verifier can run their own Bitcoin full node and Esplora instance, allowing the Prover's TEE to sync specifically against the Verifier's trusted infrastructure.
\end{itemize}

Furthermore, the Lightning Network's \texttt{channel\_reestablish} protocol (BOLT \#2) \cite{BOLT2} provides an additional peer-level safeguard. Any attempt to reconnect with a rolled-back HTLC state results in a commitment number mismatch, triggering an immediate \texttt{error} and force-close. This prevents the stale state from being fed into the TEE in the first place, complementing the enclaved chain synchronization mechanism.

\subsection{Privacy Considerations: Selective Disclosure}
\label{ssec:privacy}
Revealing exact channel balances can be detrimental to a node's competitive advantage. To address this, the Enclaved LND software can be designed to support \textbf{Selective Disclosure} mechanisms, moving beyond simple raw balance reporting.

\begin{itemize}
    \item \textbf{Range Proofs:} The enclave can be programmed to attest only that the balance exceeds a certain threshold (e.g., \enquote{Liquidity > 5 BTC}) rather than revealing the exact amount.
    \item \textbf{Zero-Knowledge Proofs (ZKP):} More advanced implementations can generate a ZK-proof inside the enclave that satisfies a solvency constraint (e.g., \texttt{Assets > Liabilities}) utilizing schemes such as Bulletproofs or zk-SNARKs. However, generating these proofs within the resource-constrained environment of an enclave introduces significant computational overhead and latency (>10s possible) \cite{Costan2016SGX}. This represents a trade-off between privacy and system performance that must be calibrated based on the use case.
\end{itemize}

\subsection{Auditor's Enhanced Verification Process}
\label{ssec:tee_verification}
The Auditor performs a multi-stage verification to build a chain of trust from the hardware to the data.

\begin{enumerate}[label=(\alph*)]
    \item \textbf{Verify Delivery via zkTLS Proof:} The Auditor first verifies the zkTLS/TLSNotary proof.
        \begin{itemize}
            \item \textbf{Action:} Use the proof and the plaintext package to confirm its validity.
            \item \textbf{Result:} Cryptographic assurance that the package containing the balance report and TEE payload was authentically served by \texttt{Node\_P}'s API (identified by its TLS certificate). This initial step prevents the Prover from fabricating the existence of a TEE attestation or replaying an old one out of context. The Auditor now trusts that the TEE payload originated from the Prover's live node at a recent time.
        \end{itemize}
    \item \textbf{Verify TEE Hardware and Quote Authenticity:} The Auditor now examines the \texttt{tee\_attestation\_payload}.
        \begin{itemize}
            \item \textbf{Action:} Send the \texttt{tee\_quote} and platform certificates to the hardware vendor's public Attestation Verification Service (e.g., Intel Attestation Service (IAS) for SGX).
            \item \textbf{Result:} The vendor's service cryptographically verifies that the quote was signed by a genuine TEE hardware key and that the TEE's security level (e.g., Trusted Computing Base, TCB) is up-to-date and not revoked. The Auditor can now trust the quote itself.
        \end{itemize}
    \item \textbf{Verify Software Integrity and Data Binding:} With a trusted quote, the Auditor inspects its contents to verify the software and bind it to the data.
        \begin{itemize}
            \item \textbf{Action (Software Version):} Extract the software measurement (MRENCLAVE) from the quote and compare it against a public list of known-good, audited measurements for trusted versions of the LND/NAM software.
            \item \textbf{Action (Data Binding):} Hash the received \texttt{balance\_report.json} and compare this hash to the \enquote{report data} hash embedded within the TEE quote.
            \item \textbf{Result:} If the MRENCLAVE matches a trusted version and the data hashes match, the Auditor is assured that the specific balance report was produced by an untampered, audited version of the node software running in a secure hardware environment.
        \end{itemize}
    \item \textbf{Final Conclusion:} Only after all three layers of checks pass does the Auditor trust the balance figures contained within the \texttt{balance\_report.json}. The trust is no longer in the Prover's honesty, but in the chain of cryptographic proofs: zkTLS for delivery, the vendor for hardware authenticity, and the TEE quote for software and data integrity.
\end{enumerate}

\subsection{Security Guarantees and Trust Assumptions}
\label{ssec:tee_guarantees}
This TEE-enhanced framework provides a robust guarantee of \textbf{software integrity}. The Auditor no longer needs to trust that the Prover is running honest software; this is now verified via hardware attestation. The trust model shifts to:

\begin{itemize}
    \item The security of the TEE hardware and its remote attestation mechanisms.
    \item The integrity of the TEE manufacturer's signing infrastructure.
    \item The quality of the audit for the published known-good software measurements (MRENCLAVE).
    \item The security of the channel through which the enclaved LND service receives live channel state. If this input can be tampered with \textit{before} entering the TEE, the guarantee is weakened, though the Enclaved Chain Synchronization mechanism described in Section~\ref{ssec:stale_state} mitigates this significantly.
\end{itemize}

\section{Comparative Analysis and Discussion}
\label{sec:comparative_analysis}
Table~\ref{tab:comparison} provides a summary comparing the three methodologies. A robust verification strategy should not be viewed as a single monolithic solution, but rather as an antifragile system composed of \enquote{Hot} and \enquote{Cold} proofs.

\subsection{Hot vs. Cold Proofs}
\begin{itemize}
    \item \textbf{Hot Proofs (Verifiable Claims):} The TEE + zkTLS framework described in this paper generates \enquote{Hot Proofs.} These are fast, efficient, and non-disruptive. They rely on hardware roots of trust and external data dependencies (like Esplora) to provide high-speed, verifiable claims of solvency suitable for regular automated checks.
    \item \textbf{Cold Proofs (The Final Arbiter):} In the event of a dispute, or if the \enquote{Hot Proof} mechanism fails (e.g., due to an API outage or TEE vulnerability), the system falls back to the \enquote{Cold Proof.} This is the on-chain settlement of funds via channel closure. It is slow and expensive, but it relies solely on the Bitcoin base layer consensus, serving as the ultimate, trustless failsafe.
\end{itemize}

\subsection{Discussion: Soundness, Liveness, and Privacy}
\label{sec:discussion}
A key trade-off in the Hot Proof model is between soundness and liveness \cite{McCorry2019PISA}.

\begin{itemize}
    \item \textbf{Soundness:} By enforcing strict chain synchronization inside the enclave, we ensure the proof is sound—it cannot lie about funds that have been slashed \cite{PoonDryja2016}.
    \item \textbf{Liveness and Privacy Leakage:} This dependency on an external API (like Esplora) introduces a liveness risk. If the API is down, the TEE cannot sync, and no proof is generated. Furthermore, utilizing a public third-party API introduces a metadata privacy leak, as the API provider can infer the node's channels based on the blocks and transactions queried by the enclave. As mitigated in Section~\ref{ssec:stale_state}, the Verifier running their own Esplora instance resolves both issues for high-stakes audits \cite{Esplora}.
\end{itemize}

\subsection{Security Considerations: Hardware Vulnerabilities}
We must also acknowledge that \enquote{Hot Proofs} are not a silver bullet. They rely on the security of the underlying TEE hardware. Historical vulnerabilities, such as Spectre, Foreshadow, and other side-channel attacks on Intel SGX, have demonstrated that it is possible for a sophisticated attacker with physical access to the machine to compromise enclave secrets. While manufacturers release microcode updates to mitigate these, the \enquote{Cold Proof} remains the necessary backstop should the hardware security class be fundamentally compromised.

\begin{table}[htbp]
\centering
\caption{Comparison of LN Reserve Verification Methods}
\label{tab:comparison}
\resizebox{\textwidth}{!}{%
\begin{tabular}{@{}lp{3.5cm}p{4.2cm}p{4.7cm}@{}}
\toprule
\textbf{Aspect} & \textbf{Auditor-Driven Probing} & \textbf{zkTLS of LND API (No TEE)} & \textbf{TEE + zkTLS (\enquote{Hot Proof})} \\ \midrule
\textbf{Trust in Node Software} & Low (verifies behavior externally) & High (Auditor must trust Prover's LND software is honest) & Low (Auditor trusts audited TEE-enclaved LND version via hardware attestation) \\ \midrule
\textbf{Proof Type} & Observational inference & Cryptographic attestation of node's \textit{statement} & TEE-attested balance + zkTLS delivery proof \\ \midrule
\textbf{Accuracy} & Bounds/estimates & Exact (as reported by LND API) & Exact (as reported by TEE-enclaved LND API) \\ \midrule
\textbf{Privacy for Target} & Low (probing is observable) & Exact balance revealed by this LND API call. Privacy depends on API design. & Tunable via Selective Disclosure (Range Proofs/ZK). \\ \midrule
\textbf{Prover Complexity} & N/A (target is passive) & Moderate (setup API, TLS cert, run zkTLS/TLSNotary client) & High (all of zkTLS + TEE setup, enclaved app dev, TEE attestation) \\ \midrule
\textbf{Auditor Complexity} & Moderate (run node, execute probing) & Moderate (verify zkTLS/TLSNotary proof, cert, JSON) & High (all of zkTLS + TEE quote verification, MRENCLAVE checks) \\ \midrule
\textbf{Resistance to Tampering} & High (external check) & Low (target can tamper LND software to lie in its API response) & High (TEE protects against software tampering) \\ \bottomrule
\end{tabular}%
}
\end{table}

\section{Industrial Relevance}
\label{sec:industrial_relevance}
The transition of the Lightning Network from a peer-to-peer payment rail to a financialized layer requires robust primitives for credit and risk assessment. The proposed \enquote{Hot Proof} framework serves as a foundational building block for the maturation of the LN industry, enabling novel commercial applications that were previously constrained by trust boundaries.

\subsection{Liquidity Marketplaces and Reputation Systems}
Current liquidity marketplaces rely on reputation scores that are either centralized or inferred from limited public data. By integrating TEE-backed verifications, these marketplaces can automate the matching of liquidity buyers and sellers with cryptographic assurance of capacity. This reduces the friction of manual auditing and enables a transition from \enquote{trust-based} reputation to \enquote{proof-based} reputation, significantly lowering the barrier to entry for new routing nodes.

\subsection{Undercollateralized Lending and Credit Markets}
A major impediment to LN adoption is the capital inefficiency of over-collateralized lending. The ability to verify a node's channel state—specifically its cash flow and solvency—in real-time without taking custody allows lenders to assess creditworthiness accurately. This framework facilitates undercollateralized lending models where credit is extended based on verifiable historical performance and current liquidity depth, mimicking traditional commercial credit lines but with cryptographic guarantees.

\subsection{Institutional Compliance and Proof of Solvency}
As institutional capital enters the Bitcoin ecosystem, compliance with financial regulations becomes paramount. Institutions require auditability to demonstrate solvency to regulators and limited partners. However, they are often unwilling to reveal trade secrets, such as specific channel partners or payment volumes. The selective disclosure capabilities of this framework (via Range Proofs or ZKPs inside the TEE) provide a mechanism for \enquote{Proof of Solvency} that satisfies regulatory audit requirements while preserving the commercial privacy necessary for institutional participation.

\section{Limitations and Future Work}
\label{sec:limitations_future_work}
The primary limitations of the proposed TEE-based framework are the current adoption rate and accessibility of TEE-enabled hardware for node operators, and the residual trust assumption regarding the security of the TEE hardware itself. Future work should focus on:

\begin{itemize}
    \item \textbf{Performance and Complexity:} Running substantial portions of the LND logic inside an enclave (like SGX) introduces I/O overhead and implementation complexity \cite{Orenbach2017Eleos}. Optimizing the boundary between untrusted host and enclave to minimize latency, especially when generating ZK-proofs, is a critical engineering challenge.
    \item \textbf{Simplifying TEE Deployment:} Developing tools and infrastructure (e.g., \enquote{Confidential Containers}) to make deploying entire applications like LND in a TEE easier for node operators \cite{CoCoCNCF}.
    \item \textbf{Securing State Input:} Researching robust methods, potentially using TEE-specific features or further cryptographic techniques, to ensure the integrity of the live channel state fed into the TEE enclave \cite{Zhang2016TownCrier}.
\end{itemize}

\section{Conclusion}
\label{sec:conclusion}
The verification of Lightning Network channel reserves requires balancing the needs of privacy, trust, and verifiability. While external probing provides an independent but imprecise measure, and simple zkTLS attestations provide authentic but not necessarily truthful statements, they lay the groundwork for more robust solutions. The primary contribution of this paper is a novel framework that leverages Trusted Execution Environments in conjunction with zkTLS to achieve robust, hardware-attested verification of LN channel balances. By proving the integrity of the balance-reporting software itself, this method provides strong guarantees against tampering by a malicious node operator, addressing a critical vulnerability in software-only attestation schemes. Although this approach introduces significant implementation complexity and new trust assumptions in TEE hardware, it represents the current state-of-the-art for high-assurance, verifiable proofs of reserves in the LN ecosystem. The continued development and standardization of this methodology will be crucial for enabling more sophisticated and trustworthy financial applications on Layer 2.

% --- BIBLIOGRAPHY ---

\end{document}